\documentclass[prd,superscriptaddress,nofootinbib,amsmath,amssymb,aps,11pt]{revtex4-2}

\usepackage{bm}
\usepackage{amsfonts}
\usepackage{latexsym}
\usepackage[latin1]{inputenc}
\usepackage{graphicx}
\usepackage{amsmath}
\usepackage{palatino}
\usepackage{mathpazo}
\usepackage[normalem]{ulem}

\usepackage{booktabs}
\usepackage{dcolumn}
\usepackage{natbib}
\usepackage[dvipsnames]{xcolor}

\begin{document}
\title{Challenges in the nonlinear evolution of unequal mass binaries in sGB gravity}

 	\author{Llibert Arest\'e Sal\'o}
	\email{llibert.arestesalo@kuleuven.be}
	\affiliation{Instituut voor Theoretische Fysica, KU Leuven. Celestijnenlaan 200D, B-3001 Leuven, Belgium}
        \affiliation{Leuven Gravity Institute, KU Leuven. Celestijnenlaan 200D, B-3001 Leuven, Belgium}

    \author{Daniela D. Doneva}
	\email{daniela.doneva@uni-tuebingen.de}
	\affiliation{Theoretical Astrophysics, Eberhard Karls University of T\"ubingen, T\"ubingen 72076, Germany}

  	\author{Katy Clough}
	\email{k.clough@qmul.ac.uk}
	\affiliation{School of Mathematical Sciences, Queen Mary University of London, Mile End Road, London, E1 4NS, United Kingdom}

 	\author{Pau Figueras}
	\email{p.figueras@qmul.ac.uk}
	\affiliation{School of Mathematical Sciences, Queen Mary University of London, Mile End Road, London, E1 4NS, United Kingdom}
	
	\author{Stoytcho S. Yazadjiev}
	\email{yazad@phys.uni-sofia.bg}
	\affiliation{Department of Theoretical Physics, Faculty of Physics, Sofia University, Sofia 1164, Bulgaria}
	\affiliation{Institute of Mathematics and Informatics, Bulgarian Academy of Sciences, Acad. G. Bonchev St. 8, Sofia 1113, Bulgaria}

\begin{abstract}
It has only recently become possible to simulate the full nonlinear dynamics of binary black holes in scalar-Gauss-Bonnet theories of gravity.
The simulations remain technically challenging and evolutions of unequal mass binaries in particular have been difficult to follow through the merger. Even when the merger is successful, accurately quantifying the physical dephasing, as opposed to contributions from transients in the initial data and gauge adjustments, remains difficult. We show the first full simulations of 2:1 and 3:1 binaries through merger, and
we discuss how specific choices in the setup affect the dephasing observed and our ability to obtain reliable results. In cases with weaker couplings, we match the expected PN value for the dephasing, whereas for larger couplings, eccentricity introduced by the initial data transients can lead to artificial deviations. 
Our work highlights the need for improvements in the initial data methods used, to ensure reliable waveforms are obtained for data analysis in beyond-GR models.
\end{abstract}
	
\maketitle

\section{Introduction}

Since the first direct detection of gravitational waves \cite{LIGOScientific:2016aoc}, observations of merging black holes have been used to test aspects of fundamental physics, putting constraints on deviations from General Relativity (GR) and the presence of new fields beyond the Standard Model \cite{LIGOScientific:2019fpa,LIGOScientific:2020tif,LIGOScientific:2021sio}. Proposed future detectors, such as LISA \cite{LISA:2017pwj}, the Einstein Telescope \cite{Punturo:2010zz}, and Cosmic Explorer \cite{ET:2019dnz,Reitze:2019iox}, will have unprecedented accuracy, allowing us to detect tens of thousands of gravitational wave events to higher redshifts and with a higher signal to noise ratio, and so further probe fundamental physics \cite{LISA:2022kgy,Perkins:2020tra,Barausse:2020rsu, Gnocchi:2019jzp,Barack:2018yly,Baker:2014zba}.
Model-independent tests of gravity look for parameterised deviations in the signals from those of GR \cite{Maggio:2022hre,Pompili:2025cdc,Krishnendu:2021fga, LIGOScientific:2021sio,Carson:2019kkh,Cornish:2011ys}, for example, by looking at changes in the coefficients of the Post-Newtonian expansion. These methods have the advantage of being fully general. However, it is not yet clear how well they would successfully capture the potential novel phenomena or correlated deviations that occur in specific theories beyond GR. Fully nonlinear modeling of the waveforms generated in such theories, using numerical relativity (NR) simulations, can help us to understand to what extent the parameterised models do a good job of capturing and identifying the new physics, and to what extent the effects may generate a systematic bias in the inferred binary parameters during data analysis.

Even within GR, binary modeling is a complex problem. Post-Newtonian, post-Minkowskian, NR and self-force approaches are still working to cover the wide range of parameter space, including effects like precession, eccentricity and higher mass ratios (e.g. some recent advances \cite{Thompson:2023ase,Ghosh:2023mhc,Wardell:2021fyy,Blanchet:2023bwj,Dhani:2024jja,Wittek:2024pis}). In modified theories of gravity, we are still at the beginning of developing accurate simulations in NR. Part of the problem is the large possible theory space for modifications, and to date, there has been a focus on theories with one extra scalar degree of freedom, mainly due to the availability of well-posed formulations in such cases \cite{Kovacs:2020pns,Kovacs:2020ywu,AresteSalo:2022hua,AresteSalo:2023mmd}. The simplest scalar-tensor theories do not generically source the scalar degree of freedom and so would not be expected to lead to deviations \cite{Healy:2011ef}. Higher order corrections change this picture and may lead to scalarised solutions \cite{Kanti:1995vq,Torii:1996yi,Sotiriou:2013qea,Doneva:2017bvd,Silva:2017uqg,Antoniou:2017acq,Kleihaus:2015aje,Kleihaus:2011tg,Cunha:2019dwb,Collodel:2019kkx,Herdeiro:2020wei,Berti:2020kgk}. Beyond initial studies in spherical symmetry \cite{Ripley:2019aqj}, fully consistent simulations including the leading parity invariant 4th order derivative corrections in an EFT framework with second order equations of motion, of which scalar-Gauss-Bonnet (sGB) is a well known example, were performed only a few years ago, first in \cite{East:2020hgw}, and then in other works
\cite{Doneva:2024ntw,Doneva:2023oww,Thaalba:2024crk,Thaalba:2024htc,Thaalba:2023fmq,Franchini:2022ukz,East:2021bqk,Corman:2024vlk,East:2022rqi,Corman:2022xqg,Corman:2024cdr,AresteSalo:2022hua,AresteSalo:2023mmd,R:2022hlf,Corman:2024vlk}. This work built on earlier studies that worked in the decoupling limit in which the scalar evolves nonlinearly but does not backreact onto the metric \cite{Okounkova:2017yby,Silva:2020omi,Elley:2022ept,Doneva:2022byd,Evstafyeva:2022rve}, which gives qualitative results but has been shown to suffer from the accumulation of secular errors \cite{Corman:2024cdr}. 

Results so far have shown a good order of magnitude agreement with the post-Newtonian expansion estimations \cite{Corman:2022xqg,Shiralilou:2020gah,Shiralilou:2021mfl,Julie:2019sab,Yagi:2011xp,Lyu:2022gdr} and although studies are at an early stage, some novel signatures have already been found \cite{Lara:2025kzj}. However, the effects are generally small in the regime of validity of the effective theory, and their exact quantification is still not at the required level of performance for data analysis, with one of the main challenges being the construction of self-consistent initial data. In \cite{Brady:2023dgu}, constraint satisfying solutions were found for binaries with an arbitrary scalar profile, but this does not lead to a state of quasistationary equilibrium. This was rectified in \cite{Nee:2024bur} in which quasistationary solutions were found, but this work only solved for the scalar profile in the decoupling limit. A full solution should therefore be achievable by combining the two approaches, and in this work we find evidence that such solutions are needed to enable the construction of reliable and accurate waveforms. More ad hoc methods, like a slow turn-on of the coupling between the scalar field and the Gauss-Bonnet invariant, whilst performing better in some cases compared to an immediate coupling turn-on, do not consistently improve the quality of the results and in large couplings can lead to eccentricity change.

Another area where more work is needed is on maintaining stable evolutions in more challenging cases beyond equal mass ratio, non-spinning cases, which constitute most of the simulations performed to date. One exception is the work of \cite{Corman:2022xqg}, in which non-equal mass cases were considered, but higher mass ratios suffered from breakdowns during merger. This problem is related to the fact that for high couplings, elliptic regions can form within the horizons of the black holes in the strong coupling regime. Whilst physically these are expected to be cured by higher order corrections that do not impact on the physics outside the horizon, they create problems in simulations if not fully excised. In our simulations we smoothly turn off the coupling within the BH horizon to avoid the formation of such regions, and this appears to mitigate the problem, allowing us to present the full merger signals in this work\footnote{We understand from discussions with the authors of \cite{Corman:2022xqg} that the same approach seems to cure the problems seen in their modified GHC formulation, so this fix does not appear to be specific to the modified puncture gauge that we use.}. There is a risk of introducing unphysical errors if the turn-off region is not fully contained within a well-resolved horizon, so again, this is a rather ad hoc technique, and better solutions would be welcome. 

    \begin{figure}
	\centering
    \includegraphics[width=1.0\linewidth]{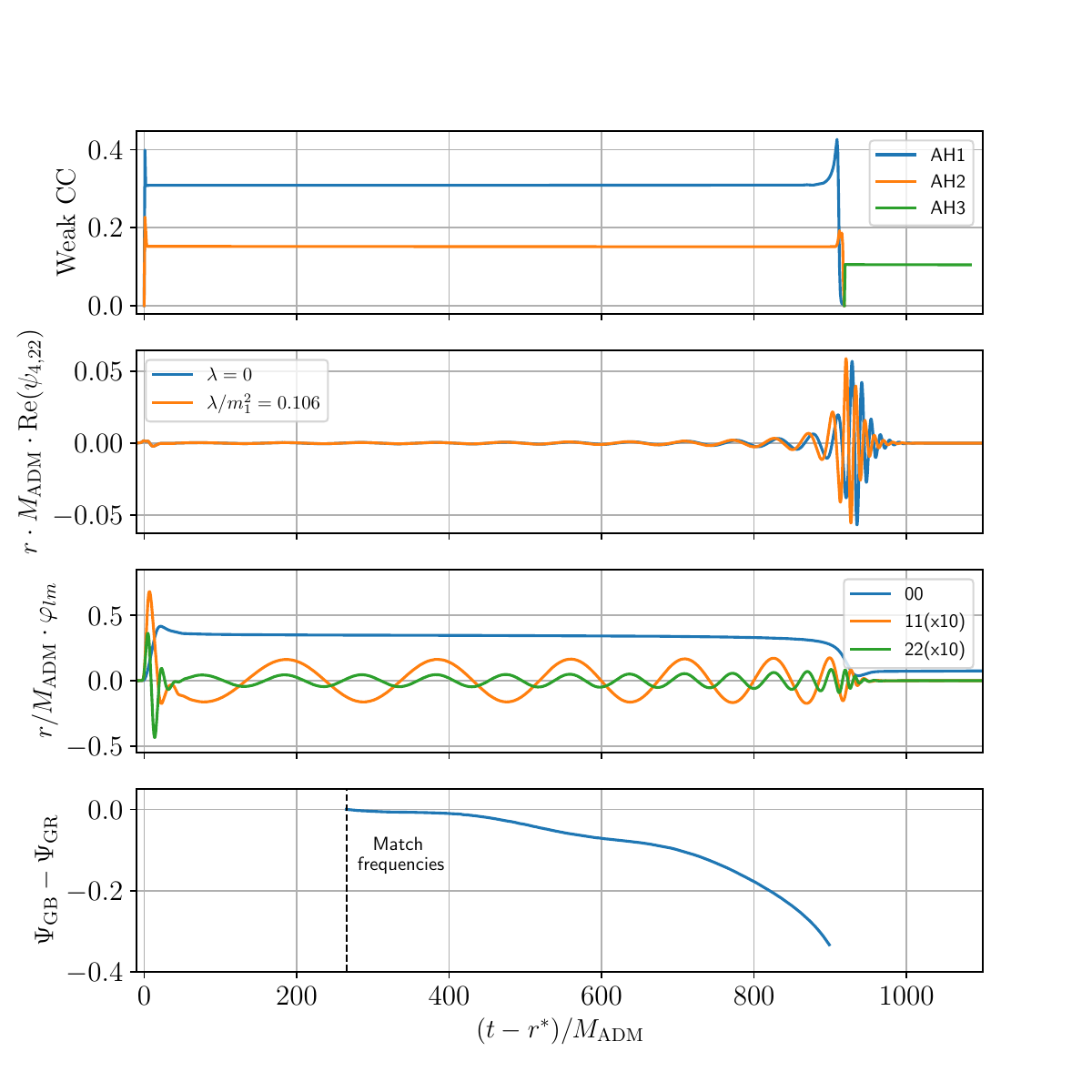}
	\caption{Overview of the $q=2$ merger. \emph{Top:} The average value of the weak coupling condition \eqref{wfc},  calculated at the apparent horizon for the $q=2$ mass ratio and $\lambda/m_1^2=0.106$. The three lines correspond to the three apparent horizons -- AH1 and AH2 are the initial two black holes, while at a later time, after the merger, a third horizon AH3 forms. \emph{Middle upper:} (2,2) mode of the real part of the Weyl scalar extracted at $r=100M$ for GR and non-GR gravity in retarded time. \emph{Middle bottom:} (0,0), (1,1) and (2,2) scalar modes extracted at $r=100M$ in retarded time. Note that we have multiplied the values of the (1,1) and (2,2) modes by 10 so that they are visible in the plot. \emph{Bottom:} Dephasing between GR and sGB waveforms.}
    \label{fig:wcc}
    \end{figure}

In this paper, we report on our work simulating non-equal mass binaries through merger.
The paper is organised as follows. In Section \ref{sec:theory} we introduce the scalar-Gauss-Bonnet theory of gravity and briefly review its $3+1$ decomposition in the modified CCZ4 formalism that we have been using, as well as the weak coupling condition that monitors when this theory is within the regime of validity of the EFT. Section \ref{sec:results} summarises the results, with a focus on the numerical set-up, how we measure the dephasing between GR and non-GR waveforms, and their impact on observables for different values of the coupling and different mass ratios. We conclude in Section \ref{sec:discussion}. 

We follow the conventions in Wald's book \cite{Wald:1984rg}. Greek letters $\mu,\nu\,\ldots$ denote spacetime indices and they run from 0 to $d$; Latin letters $i,j,\ldots$ denote indices on the spatial hypersurfaces and they run from 1 to $d$. We set $G=c=1$. 

\section{Theory and methods} \label{sec:theory}
\subsection{Scalar-Gauss-Bonnet theory of gravity}
The general form of the scalar-Gauss-Bonnet theory action is given by
	\begin{eqnarray}
		S=&&\frac{1}{16\pi}\int d^4x \sqrt{-g} 
		\Big[R - \frac{1}{2}\nabla_\mu \varphi \nabla^\mu \varphi - V(\varphi) 
		+ \lambda\,f(\varphi){\cal R}^2_{GB} \Big] ,\label{eq:quadratic}
	\end{eqnarray}
	where $R$ is the Ricci scalar,  ${\cal R}^2_{GB}$ is the Gauss-Bonnet invariant ${\cal R}^2_{GB}=R^2 - 4 R_{\mu\nu} R^{\mu\nu} + R_{\mu\nu\alpha\beta}R^{\mu\nu\alpha\beta}$, and $\varphi$ is the scalar field with a potential $V(\varphi)$. We set the potential to zero $V(\varphi)=0$, throughout the paper. The coupling strength between the scalar field and the Gauss-Bonnet invariant is controlled by $ \lambda\,f(\varphi)$, where $\lambda$ is a constant with dimensions of $length^2$. In the present paper, we focus on shift-symmetric sGB gravity with $f(\varphi)=\varphi$. We consider only $\lambda>0$ since the system is invariant under the simultaneous change of $\lambda\to-\lambda$ and $\varphi\to-\varphi$ in shift-symmetric EsGB.
    
	The variation of the action with respect to the metric and the scalar field leads to the following field equations,
	\begin{eqnarray}\label{FE}
		&&R_{\mu\nu}- \frac{1}{2}R g_{\mu\nu} + \Gamma_{\mu\nu}= \frac{1}{2}\nabla_\mu\varphi\nabla_\nu\varphi -  \frac{1}{4}g_{\mu\nu} \nabla_\alpha\varphi \nabla^\alpha\varphi - \frac{1}{2} g_{\mu\nu}V(\varphi),\\
		&&\nabla_\alpha\nabla^\alpha\varphi= \frac{dV(\varphi)}{d\varphi} -  \lambda \frac{df(\varphi)}{d\varphi} {\cal R}^2_{GB},
	\end{eqnarray}
	where   $\Gamma_{\mu\nu}$ is defined as 	
	\begin{eqnarray}
		\Gamma_{\mu\nu}&=& -\frac{1}{2}R\Omega_{\mu\nu} - \Omega_{\alpha}^{~\alpha}\left(R_{\mu\nu} - \frac{1}{2}R g_{\mu\nu}\right) 
                + 2\,R_{\alpha(\mu}\Omega^{~\alpha}_{\nu)} - g_{\mu\nu} R^{\alpha\beta}\Omega_{\alpha\beta} 
		+ \,  R^{\beta}_{\;\mu\alpha\nu}\Omega^{~\alpha}_{\beta}\,
	\end{eqnarray}  
	with 	
	\begin{eqnarray}
		\Omega_{\mu\nu}= \lambda\,\nabla_{\mu}\nabla_{\nu}f(\varphi) .
	\end{eqnarray}

\subsection{$3+1$ decomposition}
    For the numerical solution of the above system of equations, we perform a $3+1$ decomposition assuming the following form of the metric,
    \begin{eqnarray}
        ds^2=-\alpha^2dt^2+\gamma_{ij}(dx^i+\beta^idt)(dx^j+\beta^jdt)\, ,
    \end{eqnarray}
    where $\alpha$ and $\beta^i$ are the lapse function and the shift vector.

    To render the resulting system of partial differential equations hyperbolic, we adopt the modified CCZ4 formulation \cite{AresteSalo:2022hua,AresteSalo:2023mmd}, based on \cite{Kovacs:2020ywu,Kovacs:2020pns}. In this formulation, strong hyperbolicity holds as long as the system is in the weakly coupled regime. 
    The key difference from standard CCZ4 formulations is the addition of a set of terms that vanish when the constraints hold ($Z_{\mu}=0$). These additional terms supplement only the equations of motion of the metric part in \eqref{FE}  \cite{Alic:2011gg}. Specifically, the following replacement is performed,
    \begin{eqnarray}\label{mod_FE}
        R^{\mu\nu}-\tfrac{1}{2}R g^{\mu\nu}\to R^{\mu\nu}-\tfrac{1}{2}R g^{\mu\nu}+2\big(\delta_{\alpha}^{(\mu}\hat{g}^{\nu)\beta}-\tfrac{1}{2}\delta_{\alpha}^{\beta}\hat{g}^{\mu\nu}\big)\nabla_{\beta}Z^{\alpha}-\kappa_1\big[2n^{(\mu}Z^{\nu)}+\kappa_2n^{\alpha}Z_{\alpha}g^{\mu\nu} \big]\,.
    \end{eqnarray}
    Here we introduce $\hat{g}^{\mu\nu}$ and $\tilde{g}^{\mu\nu}$ as two auxiliary Lorentzian metrics that ensure that gauge modes and gauge condition violating modes propagate at distinct speeds from the physical modes, as in \cite{Kovacs:2020pns, Kovacs:2020ywu}. 
    Note that $\tilde{g}^{\mu\nu}$ does not appear explicitly in Eq. \eqref{mod_FE}, but is included in the definition of the constraints $Z^{\mu}$ (for a detailed discussion and definitions, see \cite{AresteSalo:2022hua,AresteSalo:2023mmd}).
    These auxilliary metrics are defined as
    \begin{eqnarray}\label{eq:guage_choice}
        \tilde{g}^{\mu\nu}=g^{\mu\nu}-a(x)n^{\mu}n^{\nu}\,\qquad \hat{g}^{\mu\nu}=g^{\mu\nu}-b(x)n^{\mu}n^{\nu}\, ,
    \end{eqnarray}
    where $a(x)$ and $b(x)$ are arbitrary functions such that $0<a(x)<b(x)$, and $n^{\mu}=\tfrac{1}{\alpha}(\delta_t^{\mu}-\beta^i\delta_i^{\mu})$ is the unit timelike vector normal to the $t\equiv x^0 = $const. The constraint damping terms introduced in \eqref{mod_FE} are controlled by two coefficients $\kappa_1>0$ and $\kappa_2>-\tfrac{2}{2+b(x)}$, which guarantee that the constraint violating modes are exponentially suppressed.
    The original CCZ4 formulation of \cite{Alic:2011gg} is recovered for $a(x)=b(x)=0$. In our simulations, we set $a(x)=0.2$ and $b(x)=0.4$ as in \cite{East:2021bqk}, but other values of $a(x)$ and $b(x)$ are possible and physical results should be independent of their choice \cite{Doneva:2023oww}.

    The explicit form of the $3+1$ decomposed field equations can be found in \cite{AresteSalo:2022hua,AresteSalo:2023mmd}. The versions of the $1+log$ slicing and Gamma-driver evolution equations for the lapse and shift that result in the modified puncture gauge are
    \begin{eqnarray}
        \partial_t\alpha=\beta^i\partial_i\alpha-\tfrac{2\alpha}{1+a(x)}(K-2\Theta)\,,\\
        \partial_t\beta^i=\beta^j\partial_j\beta^i+\tfrac{3}{4}\tfrac{\hat{\Gamma}^i}{1+a(x)}-\tfrac{a(x)\alpha\partial_i\alpha}{1+a(x)}\,,
    \end{eqnarray}
    where $\Theta=Z^0$, $K$ is the trace of the extrinsic curvature of the induced metric $\gamma_{ij}$, and $\tilde{\Gamma}^i=\tilde{\gamma}^{kl}\tilde{\Gamma}^i_{kl}$, with $\tilde{\Gamma}^i_{kl}$ being the Christoffel symbols associated to the conformal spatial metric $\tilde{\gamma}_{ij}\equiv\chi\gamma_{ij}$, where $\chi=\det(\gamma_{ij})^{-1/3}$.

\subsection{Weak coupling condition}

    Strong hyperbolicity in the above modified CCZ4 formulation is only guaranteed in the weak coupling regime. More importantly, beyond that regime, we cannot trust the EFT as higher order terms should play a role. We therefore require that the contributions of the Gauss-Bonnet term to the field equations, measured by the coupling $\sqrt{\lambda f'(\varphi)}$, are smaller than the two-derivative Einstein-scalar field terms. This translates into the following weak coupling condition,
    \begin{eqnarray}\label{wfc}
        \sqrt{|\lambda\,f'(\varphi)|}/L\ll 1\,.
    \end{eqnarray}
    Here $L^{-1}=\max\{
    |\nabla_{\mu}\varphi|, |\nabla_{\mu}\nabla_{\nu}\varphi|^{1/2},|{\cal R}^2_{GB}|^{1/4}\}$ is the inverse of the shortest physical length scale characterising the system, i.e., the maximum curvature scale.

    \begin{table}[htbp]
      \centering
      \setlength{\tabcolsep}{10pt}
      \caption{Parameters of the simulations presented in the papers, where $d$ is the initial distance between the two black holes, $P_x$ and $P_y$ are the initial pulses ($P_z=0$ for all simulations), and $e_{\Phi, \, max}$ is the maximum value of the eccentricity calculated using eq. \eqref{eq:ecc}. $M$ is an arbitrary code unit in the simulations, and $M_{\rm ADM}$ is the total ADM mass.}
      \begin{tabular}{|c|c|c|c|c|c|c|c|c|}
        \hline
        \hspace{-0.3cm} ($\frac{m_1}{M}$, $\frac{m_2}{M}$,$\frac{M_{\rm ADM}}{M}$) \hspace{-0.3cm} & \hspace{-0.2cm} $\lambda/m_1^2$ \hspace{-0.2cm} & \hspace{-0.2cm} $\lambda/M_{\rm ADM}^2$ \hspace{-0.2cm} & \hspace{-0.2cm} $d/M$ \hspace{-0.2cm} & \hspace{-0.2cm} $P_x/M$ \hspace{-0.2cm} & \hspace{-0.2cm} $P_y/M$ \hspace{-0.2cm} & \hspace{-0.2cm} \shortstack{\\ Slow coupling \\ turn-on} \hspace{-0.2cm} 
        & $e_{\Phi, \, max}$ \\
        \hline
        (0.5, 0.5, 1) & 0 & 0 & 11 & $9.25 \times 10^{-4}$ & 0.0897 & -- & 0.011 \\
        (0.5, 0.5, 1) & 0.04 & 0.01 & 11 & $9.25 \times 10^{-4}$ & 0.0897  & No & 0.012 \\
        (0.5, 0.5, 1) & 0.106 & 0.0265 & 11 & $9.25 \times 10^{-4}$ & 0.0897 & No & 0.019 \\
        (0.5, 0.5, 1) & 0.106 & 0.0265 & 11 & $9.25 \times 10^{-4}$ & 0.0897 & Yes & 0.010 \\
        (0.5, 1, 1.5) & 0 & 0 & 15 & 0 & 0.128  & -- & 0.062 \\
        (0.5, 1, 1.5) & 0.04 & 0.0044 & 15 & 0 & 0.128 & No & 0.062 \\
        (0.5, 1, 1.5) & 0.106 & 0.0118 & 15 & 0 & 0.128 & No & 0.060 \\
        (0.5, 1.5, 2) & 0 & 0 & 22 & $8.64 \times 10^{-4}$ & $0.136$ & -- & 0.060 \\
        (0.5, 1.5, 2) & 0.106 & 0.0066 & 22 & $8.64 \times 10^{-4}$ & $0.136$ & No & 0.060 \\
        \hline
      \end{tabular}
      \label{tab:sim_params}
    \end{table}

\subsection{Numerical set-up}
\label{subsec:num}

We have used the \texttt{GRFolres} code \cite{AresteSalo:2023hcp}, which is an extension of the NR code \texttt{GRChombo} \cite{Andrade:2021rbd}.
We define the quantities $m_1=\frac{M_{\rm ADM}}{1+q}$ and $m_2=\frac{q\,M_{\rm ADM}}{1+q}$, with $M_{\rm ADM}$ being the total ADM mass of the system. 
For all our runs, we choose initial data such that $m_1=0.5M$ (where $M$ is an arbitrary code unit), while $m_2=0.5M$, $m_2=M$ or $m_2=1.5M$ depending on the mass ratio, with $q=1$, $q=2$ or $q=3$ respectively. By fixing the length scale associated with the small black hole, we can more easily keep the same resolution for that object for all the different mass ratios. This results in a different total ADM mass for each case in code units, but in our results, we rescale the quantities so as to present them in terms of the total ADM mass $M_{\rm ADM}$ of the binary, as is conventional. The values of the coupling strength are chosen so that the scalarisation of the smaller object is (roughly) the same for each case of $q$, so they are the same in code units but different when expressed in terms of $M_{\rm ADM}$. The simulation parameters are provided in Table \ref{tab:sim_params} in code units $M$ and from this their relation to the ADM masses can be obtained. 
We use a computational domain of $L=1024M$ and $N=160$ grid points on the coarsest level (which we change to $N=128$ or $N=192$ when running at low resolution or high resolution, respectively), with $10$ levels of refinement with a refinement ratio of $2:1$, resulting in a finest resolution of $dx=M/80$, giving $\sim 80$ grid points across the horizon of the small black hole. We employ adaptive mesh refinement (AMR), using the tagging criterion explained in Section 4 of \cite{Radia:2021smk}. The CCZ4 damping parameters are fixed to $\kappa_1=1.0/M$ and $\kappa_2=-0.15$, the Kreiss-Oliger numerical dissipation coefficient is set to $\sigma = 0.5$ (see \cite{Radia:2021smk}), and finally the modified CCZ4 parameters that we employ are $a_0=0.2$ and $b_0=0.4$ (see \cite{AresteSalo:2022hua}).

The initial data has been constructed by using the \texttt{TwoPunctures} spectral solver \cite{Ansorg:2004ds}, which provides binary puncture data of Bowen-York type and is integrated into \texttt{GRChombo}. We have computed the initial momenta of the punctures at 3PN order following \cite{Bruegmann:2006ulg}. These values provide runs with at least 7 orbits before merger and relatively low eccentricity (less than $0.02$). Values are provided in Table \ref{tab:sim_params} and further information is given in Appendix \ref{app:ecc}. 
In particular, for the $q=2$ runs we have used the momentum of each puncture given to 3PN order \cite{Bruegmann:2006ulg}, while the $q=1$ and $q=3$ initial data have been taken from the MAYA catalog matching the waveform simulations MAYA0922 and MAYA0970 \cite{Ferguson:2023vta}. 

For all simulations, we use GR initial data with the scalar field initially set to zero (as in e.g. \cite{Witek:2020uzz}) and allow the BHs to scalarise dynamically. Whilst zero scalar configurations are still solutions of the constraint equations, they are not in a quasi-equilibrium state. The curvature sources the scalar field, giving rise to a non-trivial scalar configuration, which forms at the beginning of the simulation for any non-zero coupling $\lambda$. This initial hair formation disturbs the initial trajectory of the system, altering its eccentricity and affecting our ability to measure the relative dephasing correctly. 
For most of the simulations, we evolve the full sGB field equations from the beginning of the simulation. We also tested the effect of suppressing the sGB coupling $\lambda = 0$ at early times ($t \lesssim 100M$) and then slowly turning it up to the quoted value (up to $t \sim 200M$) as was done in \cite{Corman:2022xqg}. This approach has been found to reduce the induced eccentricity for certain simulations, but in general, we found it to be unreliable as a universal fix, as discussed further below.

\begin{figure}[h]
	\centering
	\includegraphics[width=1.0\linewidth]{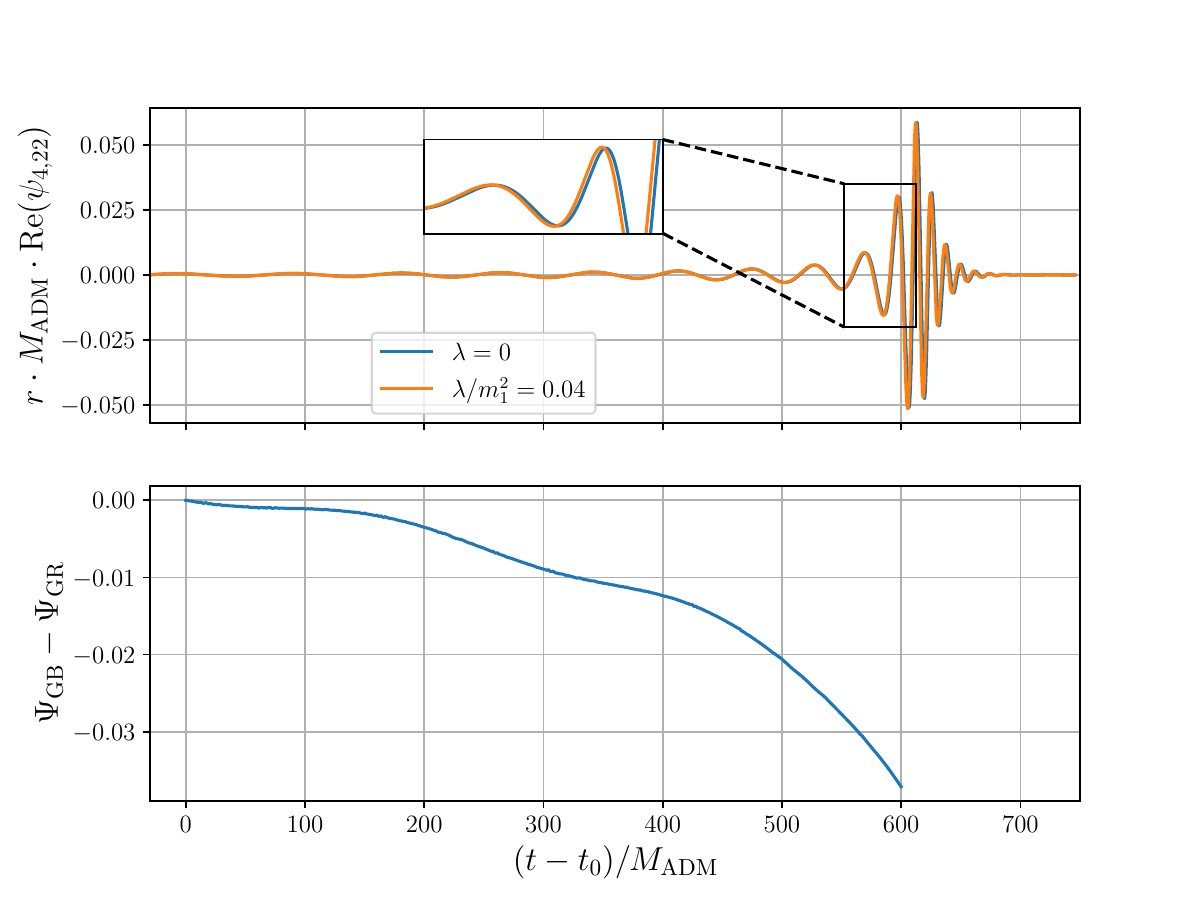}
	\caption{A $q=2$ mass ratio black hole merger in GR and shift-symmetric scalar-Gauss-Bonnet theory of gravity with low coupling $\lambda/m_1^2=0.04$. \emph{Top:} (2,2) mode of the real part of the Weyl scalar extracted at $r=150M$ for GR and sGB gravity after having aligned them at the frequency of $f_0=0.01/M_{\rm ADM}$.  \emph{Bottom:} Dephasing between GR and non-GR in the time domain. 
    }
    \label{fig:q2_low}
    \end{figure}

    \begin{figure}
	\centering
	\includegraphics[width=1.0\linewidth]{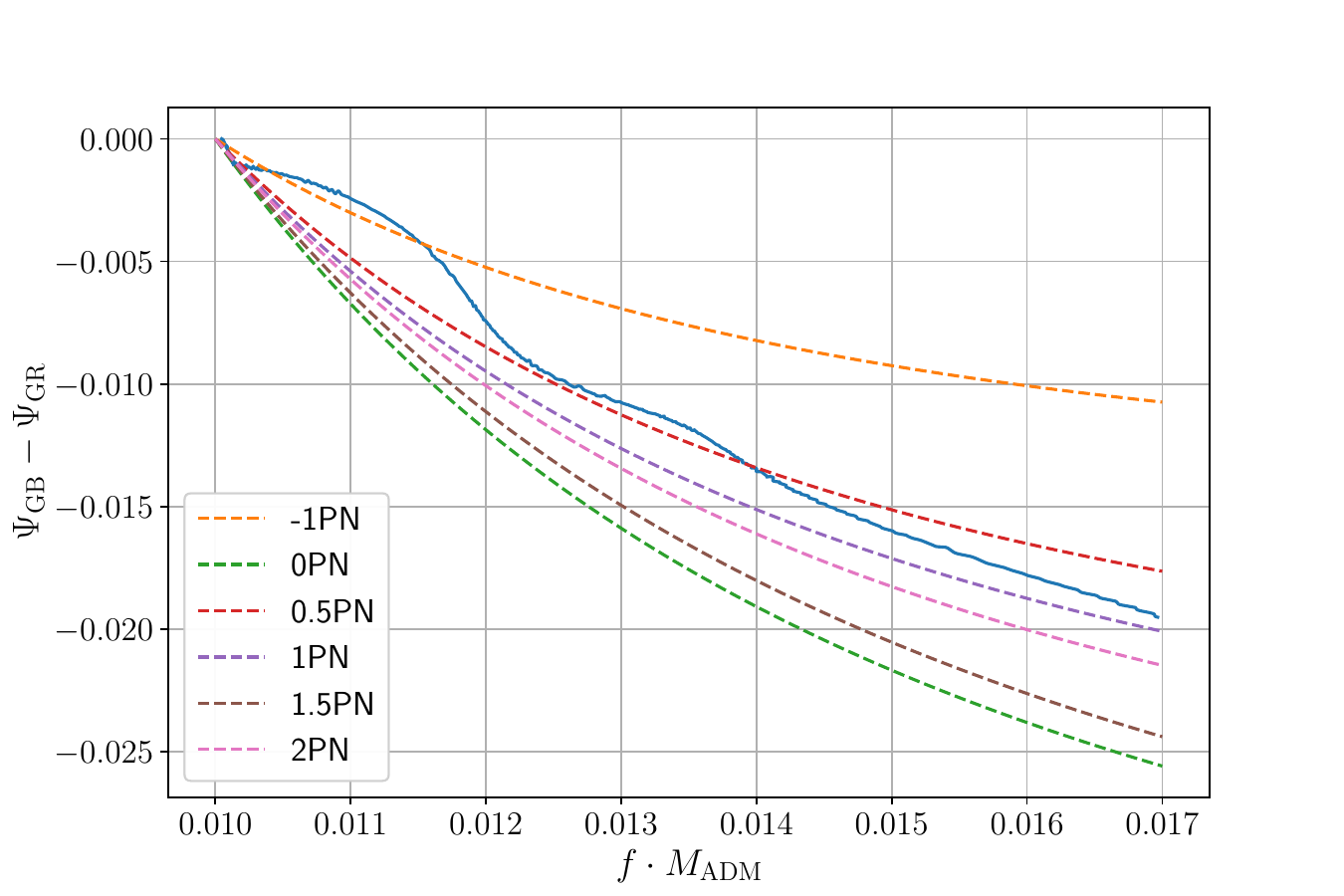}
	\caption{Dephasing between GR and sGB gravity in frequency domain for the simulation presented in Figure \ref{fig:q2_low}. We also show the post-Newtonian prediction for the dephasing calculated at different PN orders. Our results show a good agreement between the PN dephasing and our nonlinear simulations.}
    \label{fig:q2_freq}
    \end{figure}

\subsection{Measure of the dephasing between GR and non-GR waveforms}

We measure the dephasing in the gravitational waves as a function of the frequency, comparing the same initial data with zero and non zero coupling $\lambda$. We also compare this with post-Newtonian calculations, as in \cite{Corman:2022xqg}.

The frequency of the gravitational wave as a function of time can be estimated by computing the gradient of the trajectories of the punctures or the gradient of the phase of the gravitational wave. Through this, we can express the orbital phase $\Psi$, defined as half the complex phase of the (2,2) mode of the Weyl scalar, as a function of the frequency. When comparing waveforms, we align them at a given frequency. This needs to be chosen in such a way that the junk radiation from the initial data has been ejected and the eccentricity of the waveform has settled down to a small value. In most cases we have set it to $f_0=0.01/M_{\rm ADM}$, although we need to take a smaller value for cases where this frequency was attained too late in the inspiral, at which point there are too few cycles for the diagnostic to be useful. We have seen that small variations in the value of the alignment frequency do not affect the results, but choosing it to be too close to the ejection of the junk radiation or to the merger can lead to inaccuracies, which can change significantly the dephasing with respect to GR, making it too large if taken too early or too small if taken too late. The longer the inspiral, the larger the frequency band that we can use for aligning the waveforms, and so long waveforms are necessary for results to be reliable.

The post-Newtonian expressions for the orbital phase dephasing up to 2PN order that we have used can be found in Appendix B of \cite{Corman:2022xqg}. They are valid in the regime when the tensorial gravitational waves are stronger than the scalar radiation, which is the case for all our simulations. It should be noted that the PN calculations we employ lack some higher-order terms related to the non-dipolar scalar flux, which would potentially improve their accuracy. However, weak-field experimental constraints suggest that these contributions are much smaller than the 2PN GR terms \cite{Sennett:2016klh} and as in previous studies \cite{Lyu:2022gdr,Corman:2022xqg} we do not take them into account.

\section{Results}
\label{sec:results}

We present results for two values of the sGB coupling relative to the scale of the small object. In the first case $\lambda/m_1^2=0.04$ (low coupling) and we are well within the weak coupling regime. Here the phase difference with GR is smaller (but still larger than numerical error) and the eccentricity change due to the initial scalar field development is small (see Figure \ref{fig:ecc_q2} in Appendix \ref{app:ecc}). In the second case $\lambda/m_1^2=0.106$ (high coupling), which is close to the maximum allowed in observations of solar mass BHs \cite{Lyu:2022gdr,Yordanov:2024lfk} (see also Table 1 in \cite{Evstafyeva:2022rve}) 
and coincides with the higher coupling for the $q=2$ mass ratio simulations of the inspiral period in \cite{Corman:2022xqg}. For this high $\lambda/m_1$ value we are at the limit of the validity of the weak coupling condition and the change of the binary orbit due to scalar field development can be significant.

We mainly concentrate on the $q=2$ mass ratio, where we find the results to be most reliable. For higher mass ratios, namely $q=3$, we found we needed higher resolutions to reduce constraint violation, which could be due to our choice of gauge being less effective when $q$ is large. (We may need a more careful choice of gauge parameters to account for the difference in scales, as has been found to be the case in GR, see for example \cite{Muller:2010zze}.)
In this case we could still perform a stable evolution through the merger, but we believe that larger resolutions are needed to obtain more reliable results for the dephasing.

\subsection{Low coupling for $q=2$ mass ratio}
\label{subsec:12}

Here we discuss the results obtained for a $q=2$ mass ratio with GR and a low value of the coupling of $\lambda/m_1^2=0.04$.
In the top panel of Figure \ref{fig:q2_low} we compare the (2,2) mode of the Weyl scalar extracted at $r=150M$ between GR and non-GR by aligning the waveforms at $t_0$ (the time corresponding to $f_0=0.01/M_{\rm ADM}$), as discussed above. In the bottom, we show the dephasing between GR and non-GR waveforms, which we plot in the frequency domain in Figure \ref{fig:q2_freq}.
Figure \ref{fig:q2_freq} shows the post-Newtonian calculations of the dephasing with respect to GR, calculated at different post-Newtonian orders, for $\lambda/m_1^2=0.04$, $q=2$ mass ratio, and $M_{\rm ADM}=1.5M$, demonstrating that in this regime the agreement with the PN predictions of the fully nonlinear simulations is reasonably good. 

\subsection{High coupling for all mass ratios}

We now use a higher value of the coupling relative to the smaller object, $\lambda/m_1^2=0.106$, and compare this for the three different mass ratios. 

The detailed analysis of dephasing between GR and sGB performed in \cite{Corman:2022xqg}, and based on PN expansion, showed that for simulations capturing the last few orbits before the merger the highest dephasing is observed for the equal-mass system.
This is because the dipole scalar radiation (which is bigger for unequal mass ratios) is only dominant for large initial separations, whereas for smaller separations the quadrupolar radiation may drive most of the dephasing. Here we find consistent results.

We found the impact from the initial formation of hair on the eccentricity of the system to be the highest in the equal mass case $q=1$. Our results show that this was reduced, and that as a consequence we obtained a better agreement with the PN dephasing, if we slowly turned on the value of the coupling in a way similar to that done in \cite{Corman:2022xqg}, as described in Subsection \ref{subsec:num}. This can be seen in Figure \ref{fig:ecc_q1} in Appendix \ref{app:ecc}, and on the left of Figure \ref{fig:q1q2}. The slow coupling turn-on simulations reproduce the expected PN dephasing better than the immediate turn-on case. 

However, when the dephasing is lower, due to lower $\lambda$ or higher mass ratios, the eccentricity change with respect to GR is smaller and the slow turn on can give results that differ more from the PN dephasing. We tried several methods of slow turn on, and found that the results were sensitive to them, but we did not identify any simple strategy that consistently led to a smaller eccentricity change, or better agreement with the PN results for all mass ratios and values of $\lambda$. In addition, the black hole apparent horizon masses differ by around $0.1\%$ between simulations with different turn-on of the coupling, due to the different scalar radiation emitted during scalarisation. These changes could be better controlled by using initial data for scalarised black holes in quasi-equilibrium \cite{Brady:2023dgu,Nee:2024bur}, with a proper reduction of the initial eccentricity \cite{comparison}.

        \begin{figure}
	\centering
	\includegraphics[width=0.49\linewidth]{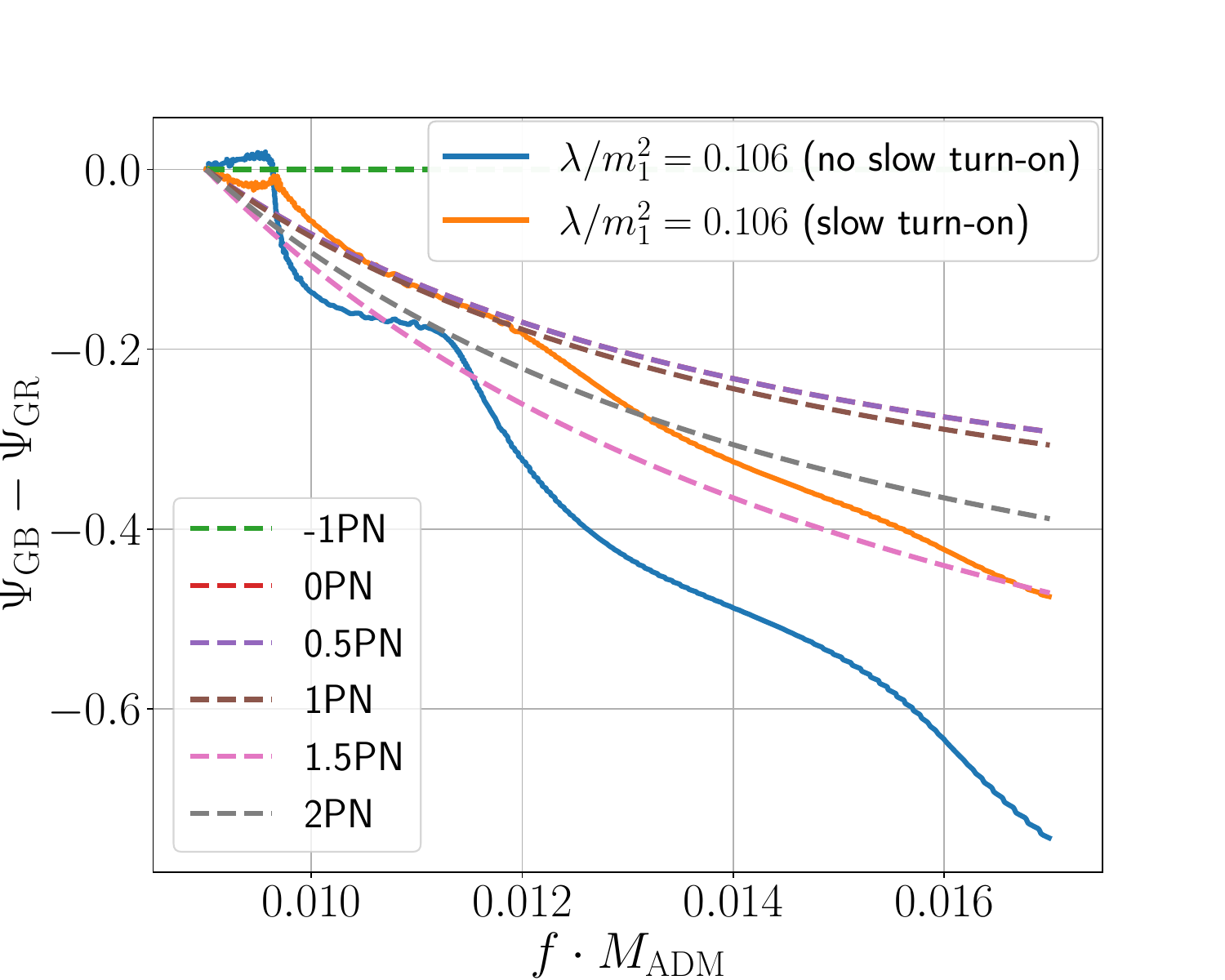}
	\includegraphics[width=0.49\linewidth]{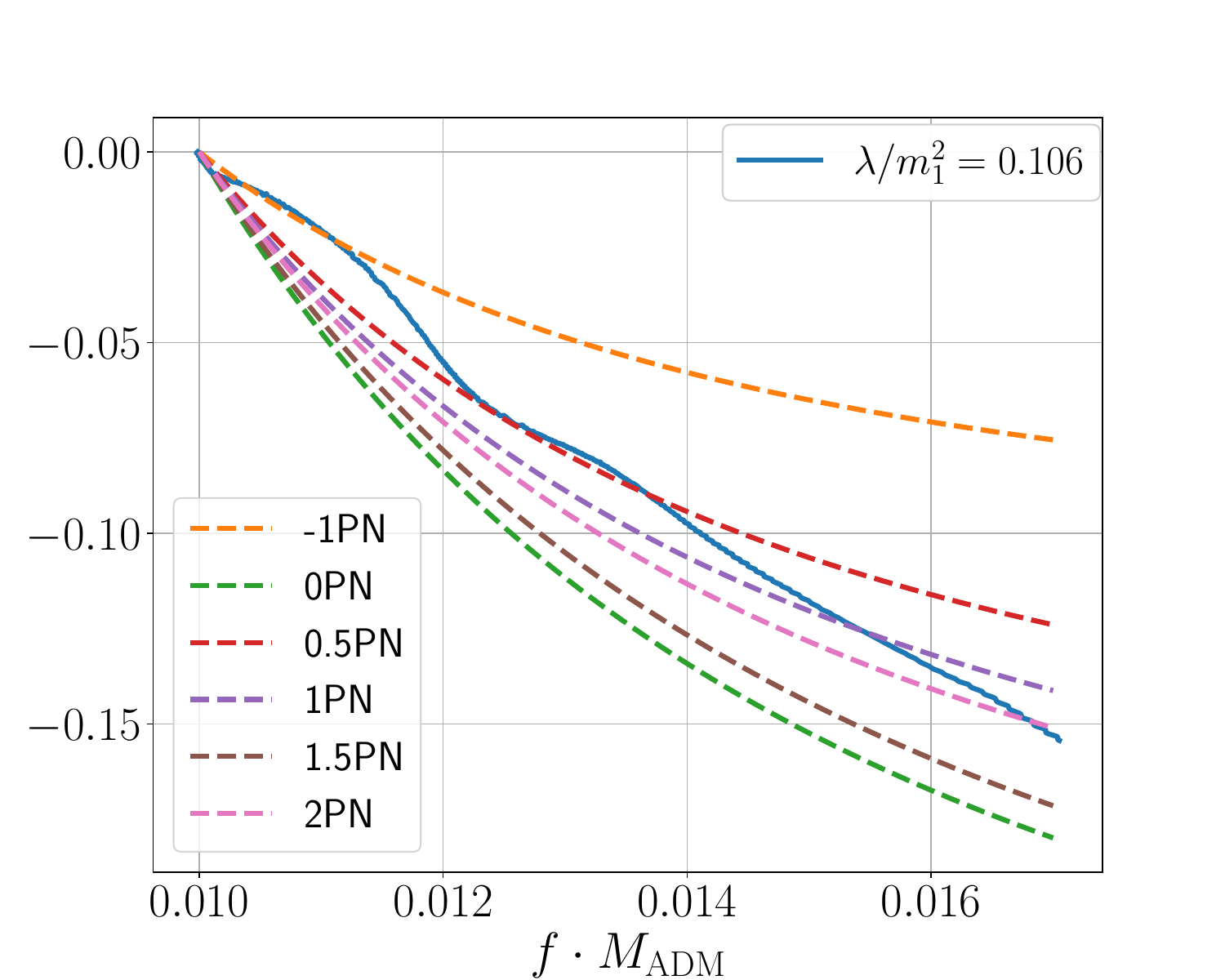}
	\caption{Dephasing between GR and sGB waveforms for $q=1$ (\emph{left}) and unequal mass $q=2$ (\emph{right}) binaries, with a high coupling of $\lambda/m_1^2=0.106$. For the equal-mass ratio case, we show both the results obtained with or without a slow turn-on of the coupling, as explained in the main text, and compare it to the PN calculations. In the first panel we see that this choice can significantly affect the results obtained, in this case making the results more consistent with PN, but in other cases the agreement was made worse.}
    \label{fig:q1q2}
    \end{figure}

\begin{figure}
	\centering
    \includegraphics[width=0.75\linewidth]{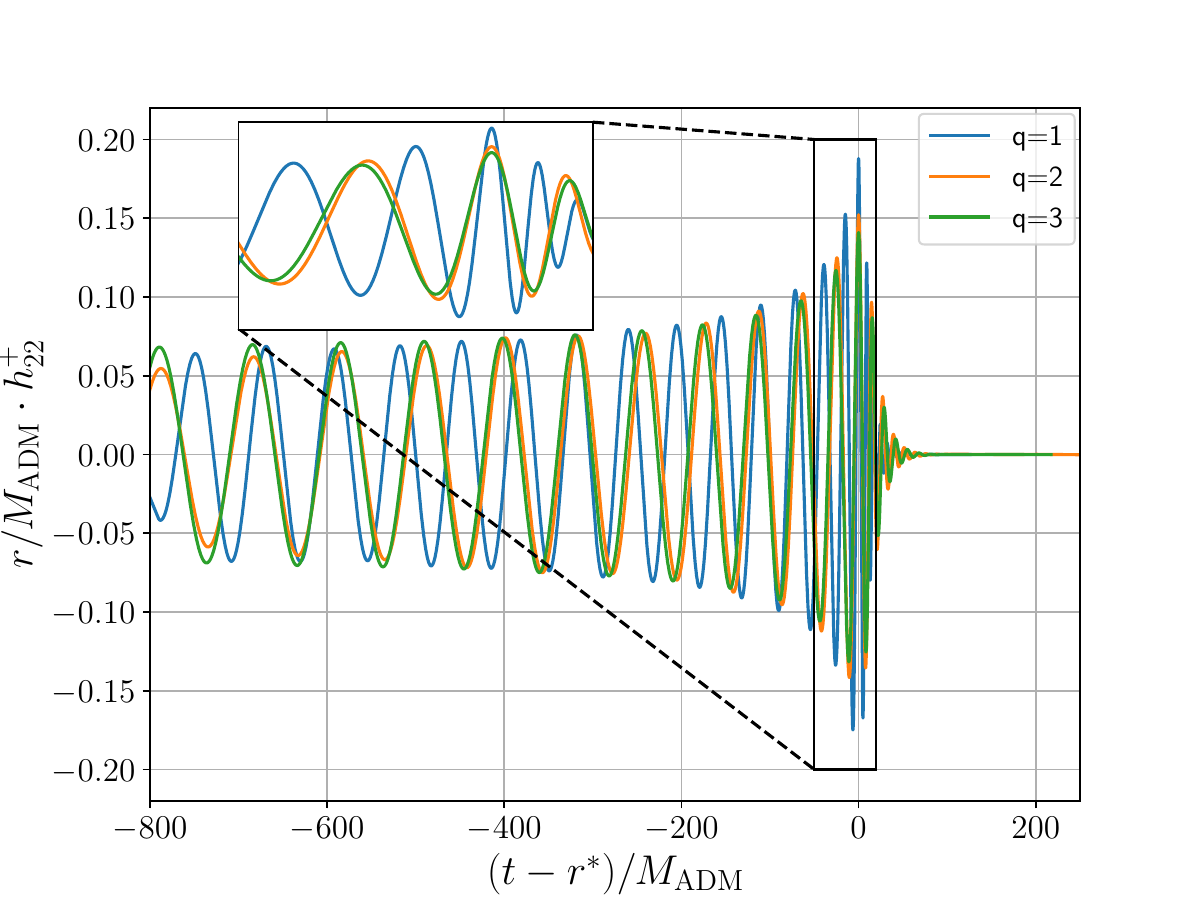}
	\caption{(2,2) mode of the \emph{plus} polarisation of the strain for the three different mass ratios considered in sGB gravity with $\lambda/m_1^2=0.106$, in retarded time $u=t-r^*$, where $r^*$ is the tortoise coordinate, extrapolated at $r\to\infty$.}
    \label{fig:q3}
    \end{figure}

In Figure \ref{fig:q3} we show the tensor waveforms obtained for all mass ratios. Whilst we are able to evolve the $q=3$ case successfully through the merger, the constraint violations are higher than in the other cases as discussed above. The dephasing is also closer to the numerical error, and so we do not analyse it in detail.

\begin{figure}
	\centering
    \includegraphics[width=0.75\linewidth]{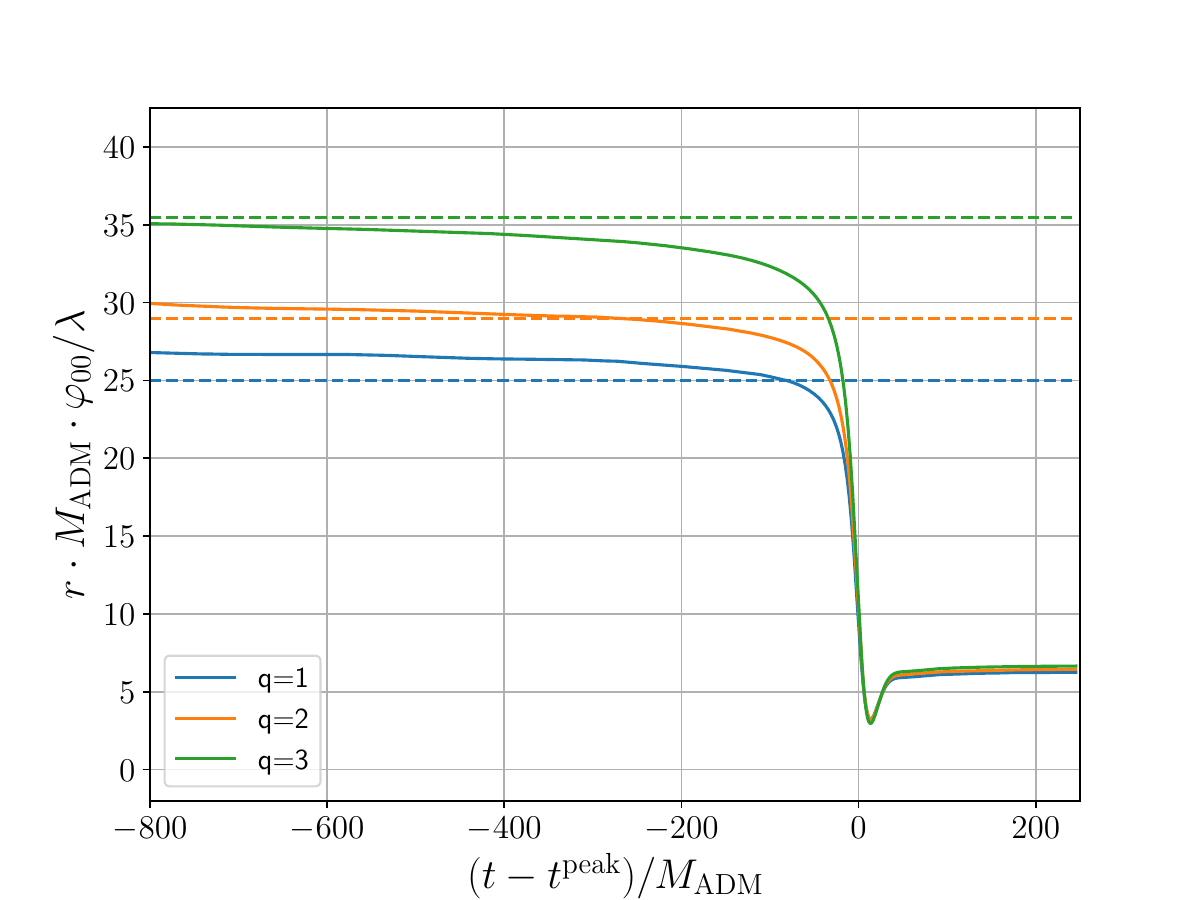}
	\caption{The (0,0) mode of the scalar waves (scalar charge) for the three different mass ratios considered in shift-symmetric sGB gravity with $\lambda/m_1^2=0.106$, aligned at the time of the merger, extracted at $r=100M$. The dotted lines show the value of the amplitude of the merger multiplied by the analytical factor $\frac{M_{\rm ADM}^2}{m_1m_2}$, which should approximate the relative amplitudes in the inspiral, with the approximation becoming better for larger $q$ as expected.}
    \label{fig:scalar}
    \end{figure}

\begin{figure}
	\centering
    \includegraphics[width=0.49\linewidth]{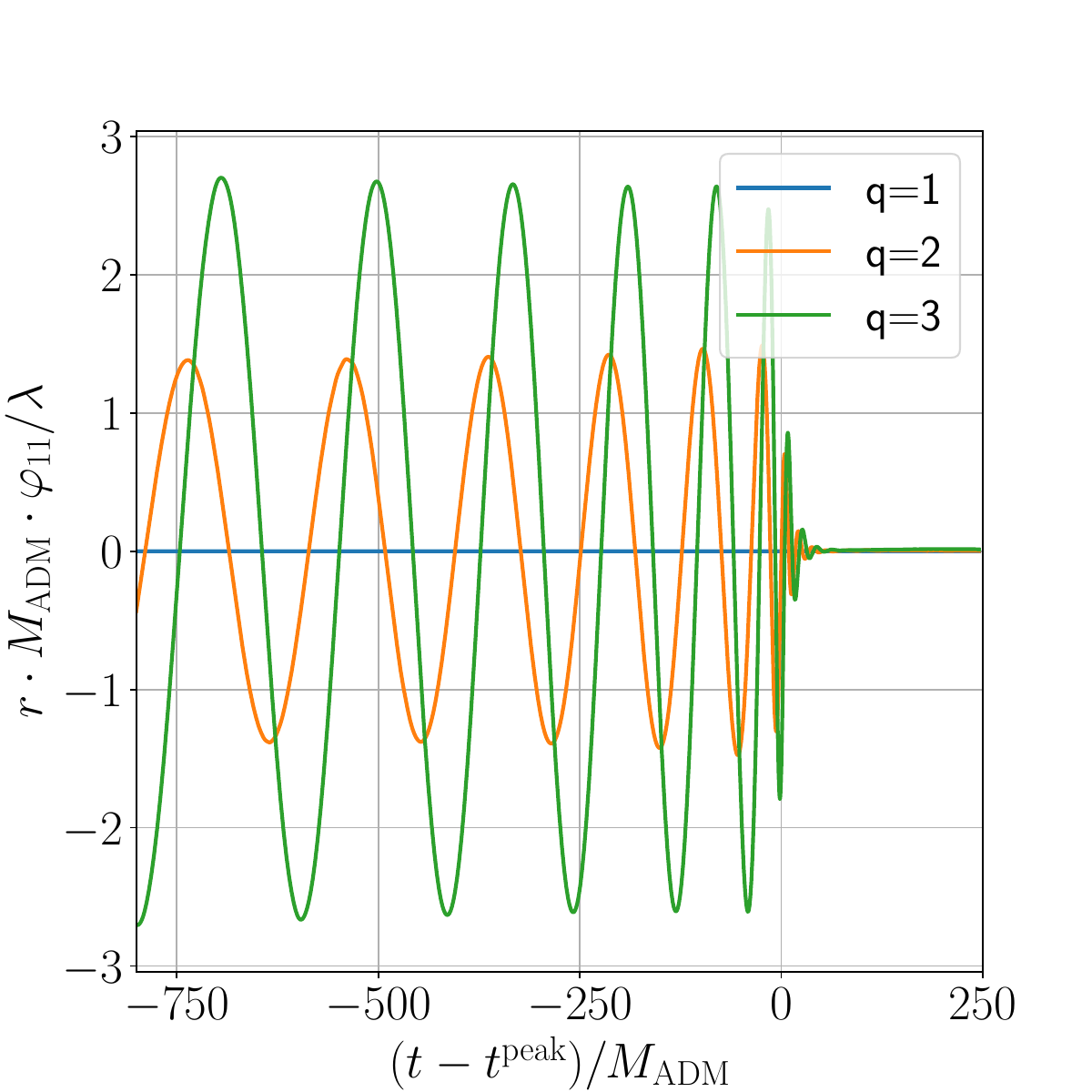}
    \includegraphics[width=0.49\linewidth]{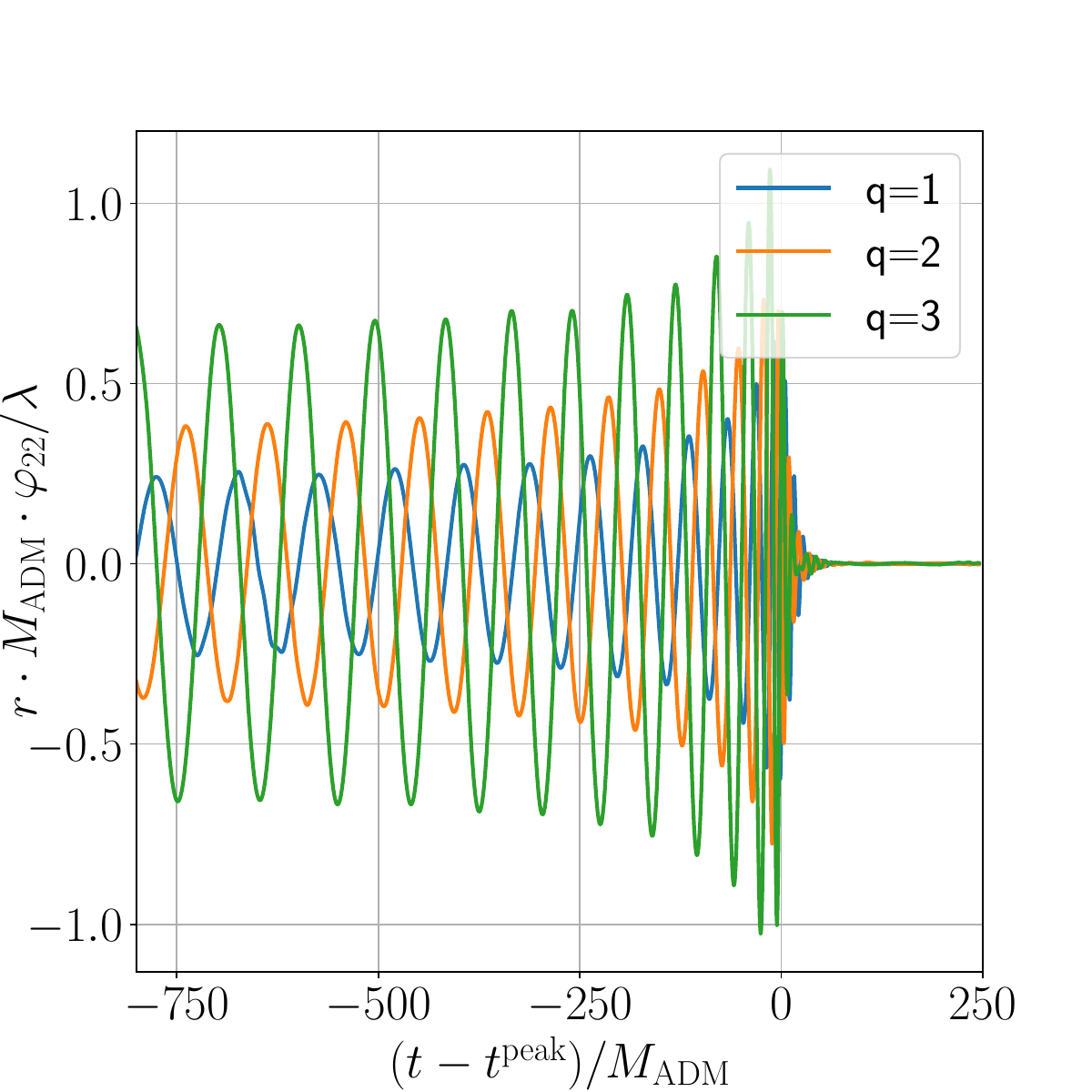}
	\caption{(1,1) and (2,2) mode of the scalar waves for the three different mass ratios considered in shift-symmetric sGB gravity with $\lambda/m_1^2=0.106$, aligned at the time of the merger.}
    \label{fig:scalar2}
    \end{figure}

    We also investigated the behavior of the dominant modes of the scalar waves for the three different mass ratios. Since the scalar field is proportional to the value of the coupling $\lambda$ when backreaction is ignored, we consider the quantity $M_{\rm ADM}^2\varphi/\lambda$ (the scalar modes are multiplied by $r/M_{\rm ADM}$ so we display $r M_{\rm ADM}\varphi/\lambda$ in our plots).
    Figure \ref{fig:scalar} displays the (0,0) mode of the scalar waves (namely the scalar charge), whose pre-merger amplitude is larger for higher values of the mass ratio, while the final amplitude is approximately independent of $q$. This behaviour is as expected, since we should find that approximately $\varphi\propto \frac{\lambda}{r}(\frac{1}{m_1} + \frac{1}{m_2})$ during the inspiral, while $\varphi\propto\frac{\lambda}{rM_{\rm ADM}}$ after merger and, indeed, one can verify that the ratio between the initial and final amplitudes is approximately given by $\frac{M_{\rm ADM}^2}{m_1m_2}$ \cite{Witek:2018dmd}, as shown by the horizontal lines in Figure \ref{fig:scalar}. This agreement is better for more unequal systems where the above approximation holds better -- the scalar charge decreases with increasing black hole mass and for more unequal systems the scalar charge of the small black hole dominates. 
    In Figure \ref{fig:scalar2} we show the evolution of the (1,1) and (2,2) modes of the scalar waves. As expected, the amplitude increases with the mass ratio and the (1,1) mode vanishes for equal-mass ratio because there is no dipolar radiation.

We end this subsection by discussing the weak coupling condition defined by Eq. \eqref{wfc}. In the top panel of Fig. \ref{fig:wcc}, we present the average value calculated at the apparent horizon for the $q=2$ case, which we show together with its corresponding gravitational and scalar waves, as well as the dephasing, in the bottom panels in order to have a complete picture. 
At $t=0$, the weak coupling condition vanishes because we start with zero scalar field initial data, and for this simulation, there is no slow turn-on of the coupling. As a result, the weak coupling condition grows rapidly and settles to a nearly constant value until shortly before the merger. During the last orbit before the plunge, the weak coupling condition grows again until the apparent horizon of the new black hole (AH3) is formed. The weak coupling condition is maximal for the smallest black hole apparent horizon (AH1) and minimal for the newly formed, massive, black hole (AH3). We see that for this value of $\lambda/m_1^2$, the system is close to leaving the regime in which it is weakly coupled.

    \subsection{Ringdown}

    In this last subsection, we briefly study the ringdown of the gravitational waveform for the cases in which we have observed the highest dephasing from GR and, therefore, we also expect the highest deviation in the values of the quasinormal modes.

    \begin{figure}[h]
	\centering
     \includegraphics[width=0.49\linewidth]{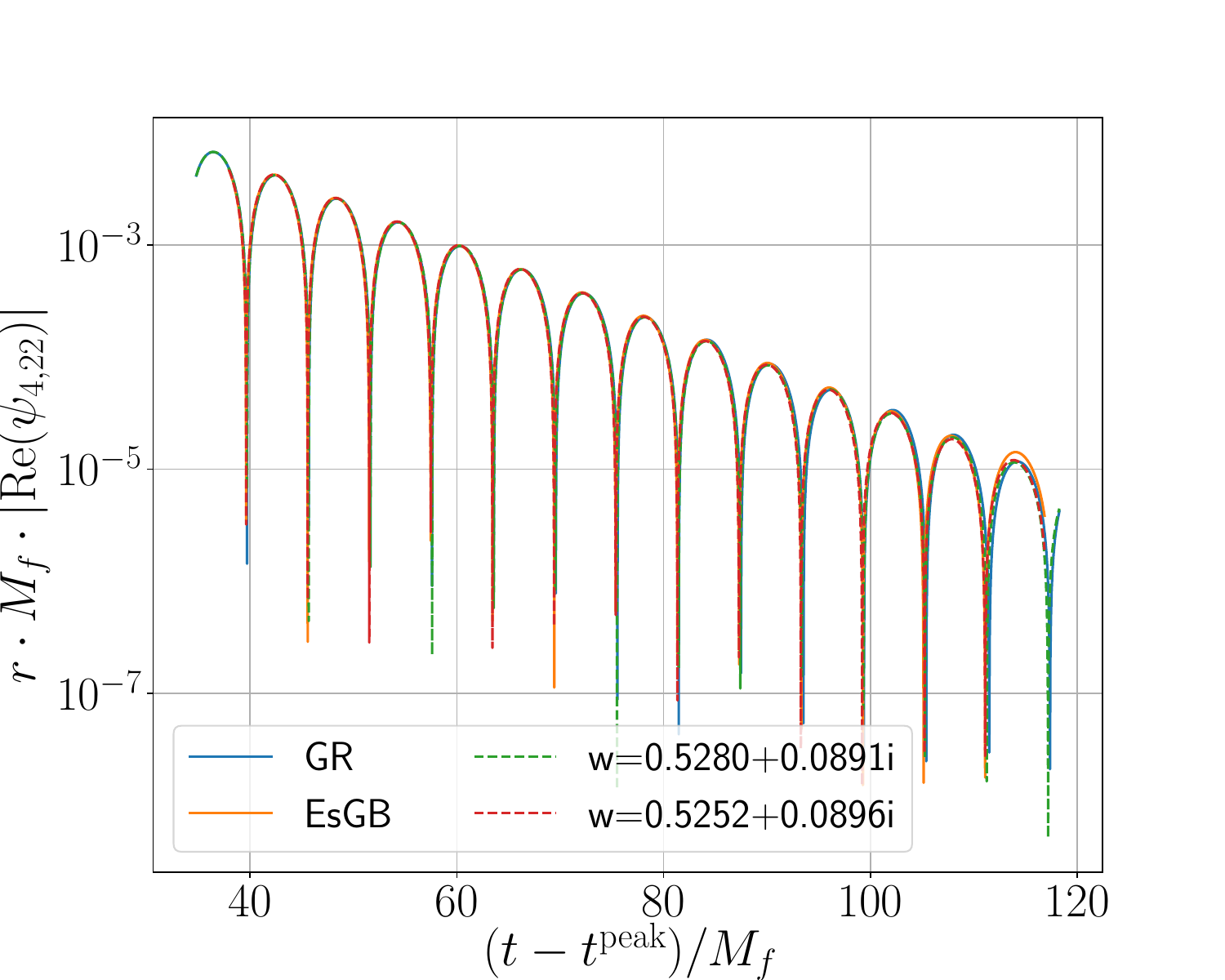}
    \includegraphics[width=0.49\linewidth]{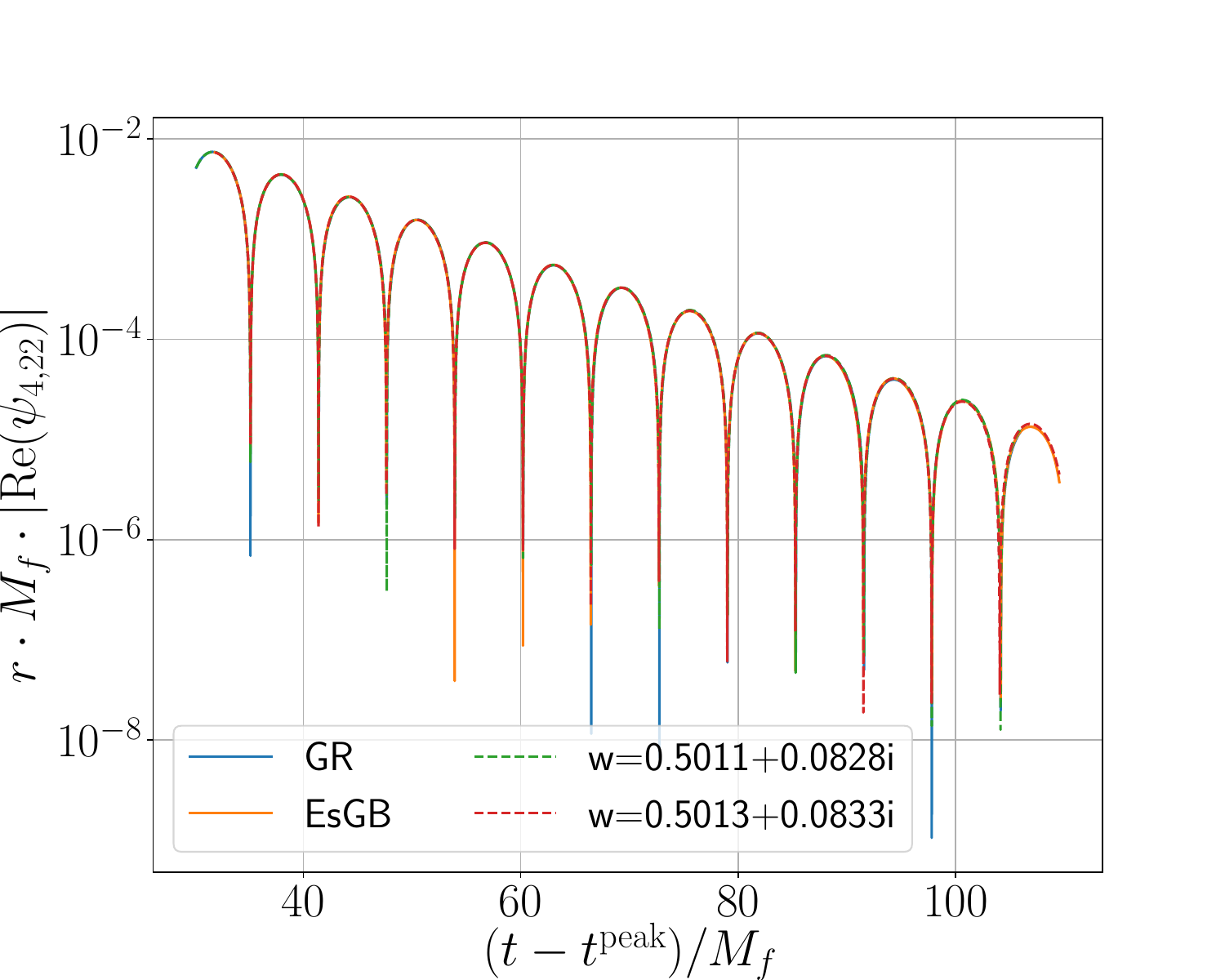}
	\caption{Ringdown of the (2,2) mode of the Weyl scalar, for equal mass (\emph{left}) and q=2 mass ratio (\emph{right}), both in GR and EsGB with $\lambda/m_1^2=0.106$. We display in dashed lines for each case the fitting with a single mode. Whilst consistent with the expected values, the dephasing is comparable to our numerical errors, and more work would be needed to obtain precise enough results.}
    \label{fig:qnms}
    \end{figure}

    In Figure \ref{fig:qnms} we show on a logarithmic scale the ringdown of the gravitational waveforms obtained for both $q=1$ and $q=2$ binaries for GR and EsGB with the high value of the coupling considered in the previous subsection, $\lambda/m_1^2=0.106$. We fit both the GR and EsGB ringdown with a single mode, which lies very close in both cases to the GR fundamental (2,2) mode. In the $q=1$ case in GR we expect $M_fw=0.5291+i\,0.0884$ for the final values of the mass and spin of the remnant found in the simulation ($M_f = 0.9510M$ and $a_f = 0.6525M$), and $M_fw=0.5045+i\,0.0801$ in the $q=2$ case (where $M_f=1.4510M$ and $a_f=0.9005M$).
    
    As expected for this value of the coupling, the difference between GR and EsGB is very small and less than our numerical error \cite{Chung:2024vaf, Khoo:2024agm}. The highest shift is expected for the equal-mass case, where the theoretical value in EsGB for the coupling used (Table V in \cite{Chung:2024vaf}) is $0.5286+0.0886i$. For the $q=2$ case, the value is even closer to GR, namely $0.5044+0.0801i$ \footnote{Note that isospectrality is broken in sGB and, thus, the frequencies differ between the polar and axial sectors. Here we have just considered the frequencies of the polar sector, which we expect to be dominant, but note that those of the axial sector are very similar for the values of the coupling that we have used.}. In both cases we find that the values shift in the right direction, but do not recover the exact values.

    A detailed analysis of the ringdown signal, including study of the mode amplitudes and careful treatment of the starting time of the post-merger extraction, requires higher resolution and more sophisticated data analysis techniques \cite{Cheung:2023vki}, and is left to future work.

    \section{Discussion}
    \label{sec:discussion}

    In this work, we have studied binary mergers in the shift symmetric EsGB theory using simulations that capture the full nonlinearities in the scalar and metric sectors, whilst remaining in the valid range for the EFT. Our work includes the merger phase for unequal mass cases, which is particularly challenging for the newly developed stable gauges for large couplings.
    
    Whilst our results are broadly consistent with expectations and recover the expected behaviour and scalings in all the regimes studied, they highlight the need for more work in order to obtain precision results for data analysis. 
    In particular, we find that the exact dephasing of the binary is sensitive to the initial data and the dynamical development of the scalar hair. Whilst a slow turn on of the coupling can help in some cases, the combination of gauge and scalarisation effects in the initial data generally lead to unwanted dynamics that increase the eccentricity and lead to unreliable results. This makes the dephasing results more sensitive to the frequency at which the GR and non GR signals are aligned, and can result in significant deviations -- from results that agree with PN predictions to ones that seem to deviate significantly from them. For that reason, a lot of care needs to be taken to reduce initial eccentricity and judiciously align the signals, which makes obtaining robust results a challenge. 
    
    Our results indicate that codes that simulate beyond-GR models will need to implement initial data that imposes a quasi-equilibrium state of the scalar field, as achieved in \cite{Nee:2024bur}, whilst fully solving the constraints, as in \cite{Brady:2023dgu}, in combination with eccentricity reduction on the initial data to obtain more consistent parameters \cite{Pfeiffer:2007yz}. With this they will be better placed to produce waveforms to test beyond-GR data analysis pipelines.

    \section*{Acknowledgements}
    We thank Josu Aurrekoetxea, Maxence Corman, Will East, Tamara Evstafyeva and Miren Radia for useful discussions and Lorenzo Rossi for initial work on the unequal mass cases. We thank the entire \texttt{GRTL} Collaboration\footnote{\texttt{www.grtlcollaboration.org}} for their support and code development work.
    This study is in part financed by the European Union-NextGenerationEU, through the National Recovery and Resilience Plan of the Republic of Bulgaria, project No. BG-RRP-2.004-0008-C01. DD acknowledges financial support via an Emmy Noether Research Group funded by the German Research Foundation (DFG) under grant no. DO 1771/1-1. KC is supported by an STFC Ernest Rutherford fellowship, project reference ST/V003240/1. PF and KC are supported by an STFC Research Grant ST/X000931/1 (Astronomy at Queen Mary 2023-2026). During the early stages of the work, LAS was supported by an LMS Early Career Fellowship. LAS is partly funded by IBOF/21/084. We acknowledge Discoverer PetaSC and EuroHPC JU for awarding this project access to Discoverer supercomputer resources. This work also used the DiRAC@Durham facility managed by the Institute for Computational Cosmology on behalf of the STFC DiRAC HPC Facility (www.dirac.ac.uk). The equipment was funded by BEIS capital funding via STFC capital grants ST/P002293/1, ST/R002371/1 and ST/S002502/1, Durham University and STFC operations grant ST/R000832/1. DiRAC is part of the National e-Infrastructure.

    \appendix
    \section{Convergence}
    \label{app:conv}
   
\begin{figure}
	\centering
        \includegraphics[width=0.75\linewidth]{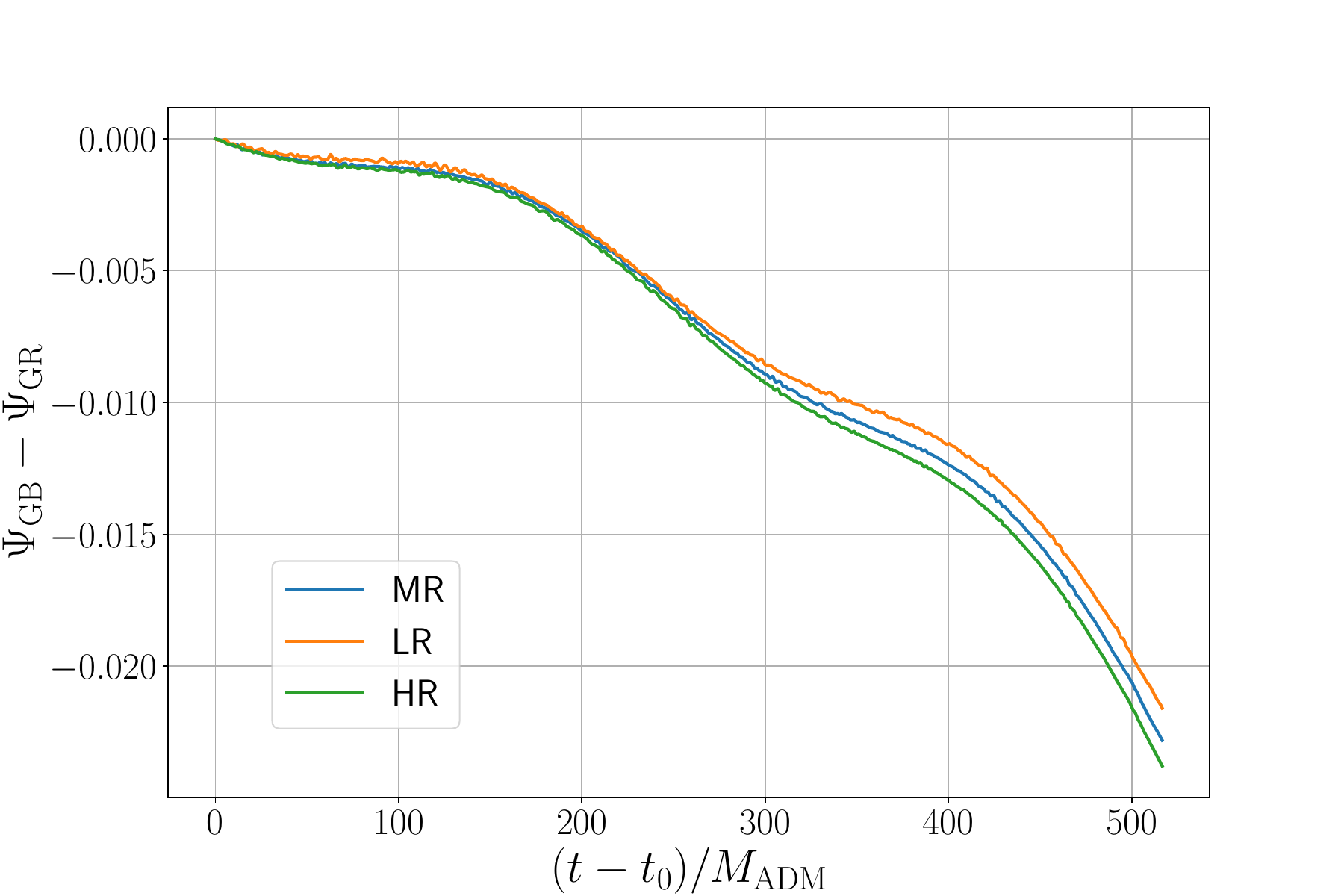}
	\caption{Dephasing between GR and non-GR (with $\lambda/m_1^2=0.04$) for a $1:2$ binary in the time domain for low-resolution (LR), medium-resolution (MR), and high-resolution (HR) simulation.}
    \label{fig:conv_dephasing}
    \end{figure}

    In this section, we present the convergence test for the orbital phase $\Psi$ defined in this paper for the $q=2$ binary black hole merger in sGB gravity with $\lambda/m_1^2=0.04$ discussed in Section \ref{subsec:12}. We show in Figure \ref{fig:conv_dephasing} the dephasing with respect to GR from Figure \ref{fig:q2_low} for the three different resolutions we have considered, namely $N=128$ (low), $N=160$ (medium), and $N=192$ (high).
    
            \begin{figure}
	\centering
    \includegraphics[width=0.6\linewidth]{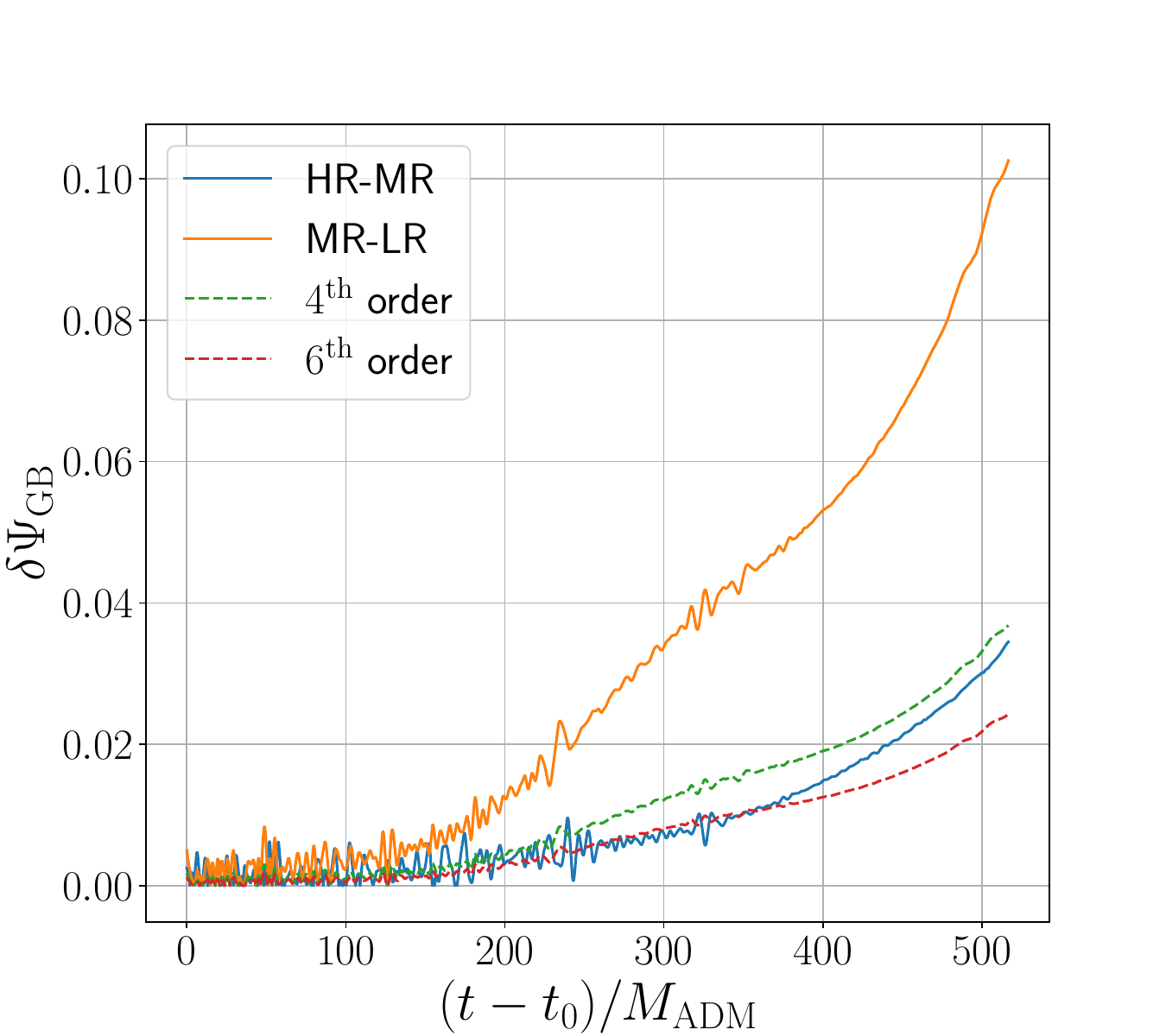}
	\caption{Differences of the orbital phase $\Psi$ across three different resolutions for a $1:2$ binary in sGB gravity with $\lambda/m_1^2=0.04$. The difference between the low and the medium resolution runs has been rescaled by the convergence factor $Q_n=\frac{h_{MR}-h_{HR}}{h_{LR}-h_{MR}}$ (with $h_i$ being the grid spacing for the different resolutions), assuming fourth and sixth order convergence.}
    \label{fig:conv}
    \end{figure}
    
    Figure \ref{fig:conv} explores further the convergence, demonstrating that during the inspiral and merger phases of the binary, the convergence order of $\Psi$ is around six and it reduces to four as we approach the merger. This is consistent with the order of the finite difference stencils used. The results of the convergence analysis presented here indicate that our simulations are stable and in the convergent regime.

    \section{Eccentricity}   \label{app:ecc}
    Following \cite{Mroue:2010re}, a convenient way to define the orbital eccentricity is through the gravitational wave phase. For that purpose, we extract the phase and amplitude of the $(l,m)=(2,2)$ component of $\psi_4$, namely $\Psi$, in the following way,
    \begin{equation}
        r \times \psi_{4,22} = A_{22}(t,r) e^{-i \Phi_{22}}.
    \end{equation}
    We fit the resulting phase $\Phi_{22}$ with a 7th order polynomial, to obtain $\Phi_{\rm fit}$. Thus, the eccentricity is defined as
    \begin{equation}\label{eq:ecc}
        e_\Phi (t) = \frac{\Phi_{22} - \Phi_{\rm fit}}{4}.
    \end{equation}
    Ideally, $e_\Phi$ will be a sinusoidal function of $t$ and the orbital eccentricity is defined as the maximum absolute value of this function. The pure sinusoidal dependence is slightly spoiled, though, by different factors, including junk radiation during the initial gauge settling and the limited timeframe during which we measure the eccentricity. In the sGB case, the eccentricity is further affected by the dynamical scalarisation of the two black holes.
   
    \begin{figure}
	\centering
        \includegraphics[width=0.49\linewidth]{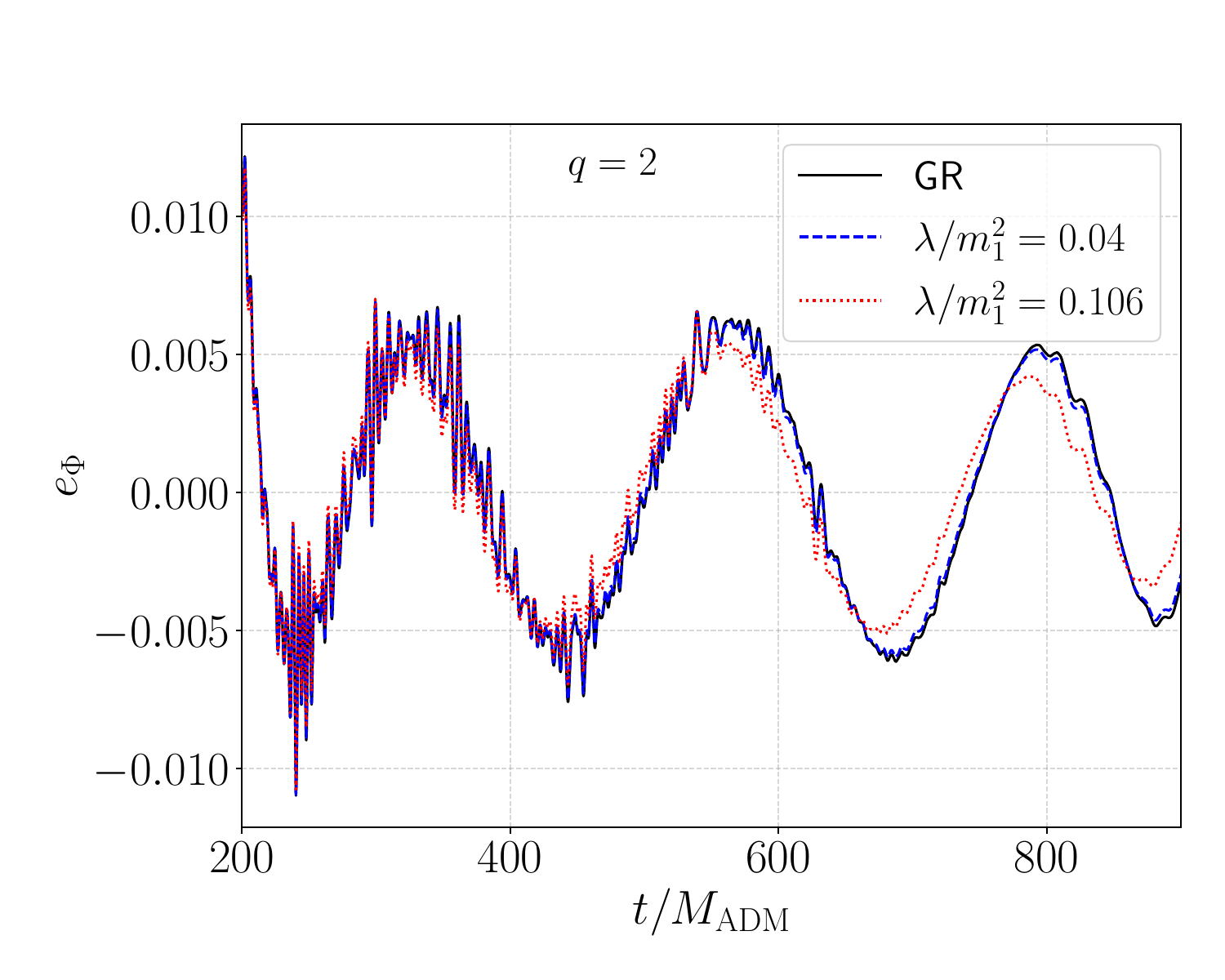}
        \includegraphics[width=0.49\linewidth]{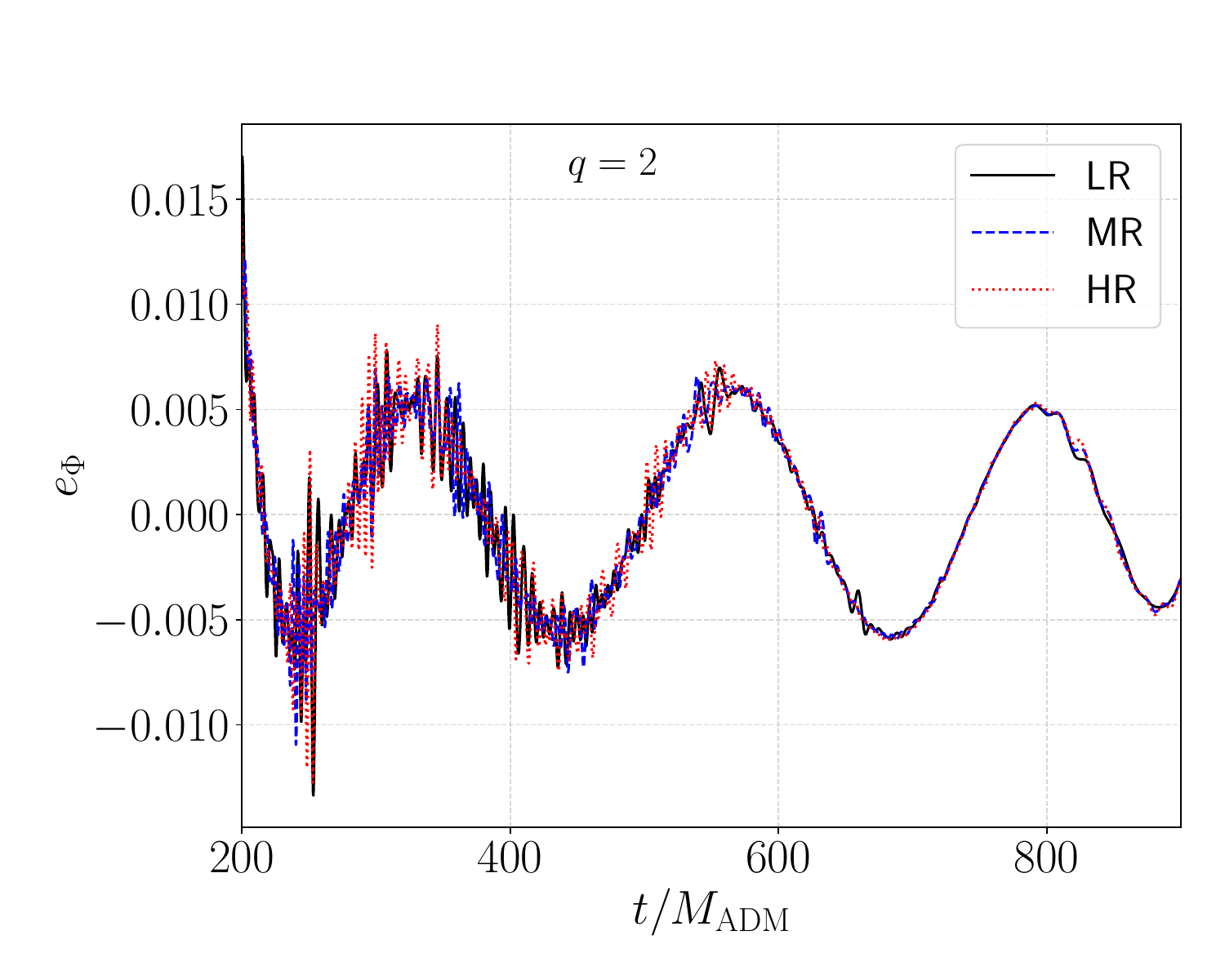}
	\caption{The calculated eccentricity $e_\Phi$ for $q=2$ mass ratio black hole mergers in GR and shift-symmetric scalar-Gauss-Bonnet theory of gravity. \emph{Left:} Results for GR and $\lambda/m_1^2=\{0.04, 0.106\}$. \emph{Right:} Results for the three different resolutions with $\lambda/m_1^2=0.04$. The turn-on of the coupling is immediate, i.e., there is no slow turn-on. }
    \label{fig:ecc_q2}
    \end{figure}
    
    The calculated eccentricity $e_\Phi$ for the $q=2$ mass ratio simulations is presented in Fig. \ref{fig:ecc_q2} and it reaches up to roughly $0.06$. As one can see in the right panel, $e_\Phi$ changes only slightly for different resolutions. Therefore, the resulting eccentricity is mainly due to the imperfect initial data. With the increase of the parameter $\lambda$, though, larger differences with GR are observed. While for $\lambda/m_1^2=0.04$ the $e_\Phi(t)$ dependence is almost indistinguishable from GR, in the $\lambda/m_1^2=0.106$ simulations, the deviations are already noticeable. Nevertheless, the maximum deviation is of the order of $10-20\%$. 

           \begin{figure}
	\centering
        \includegraphics[width=0.8\linewidth]{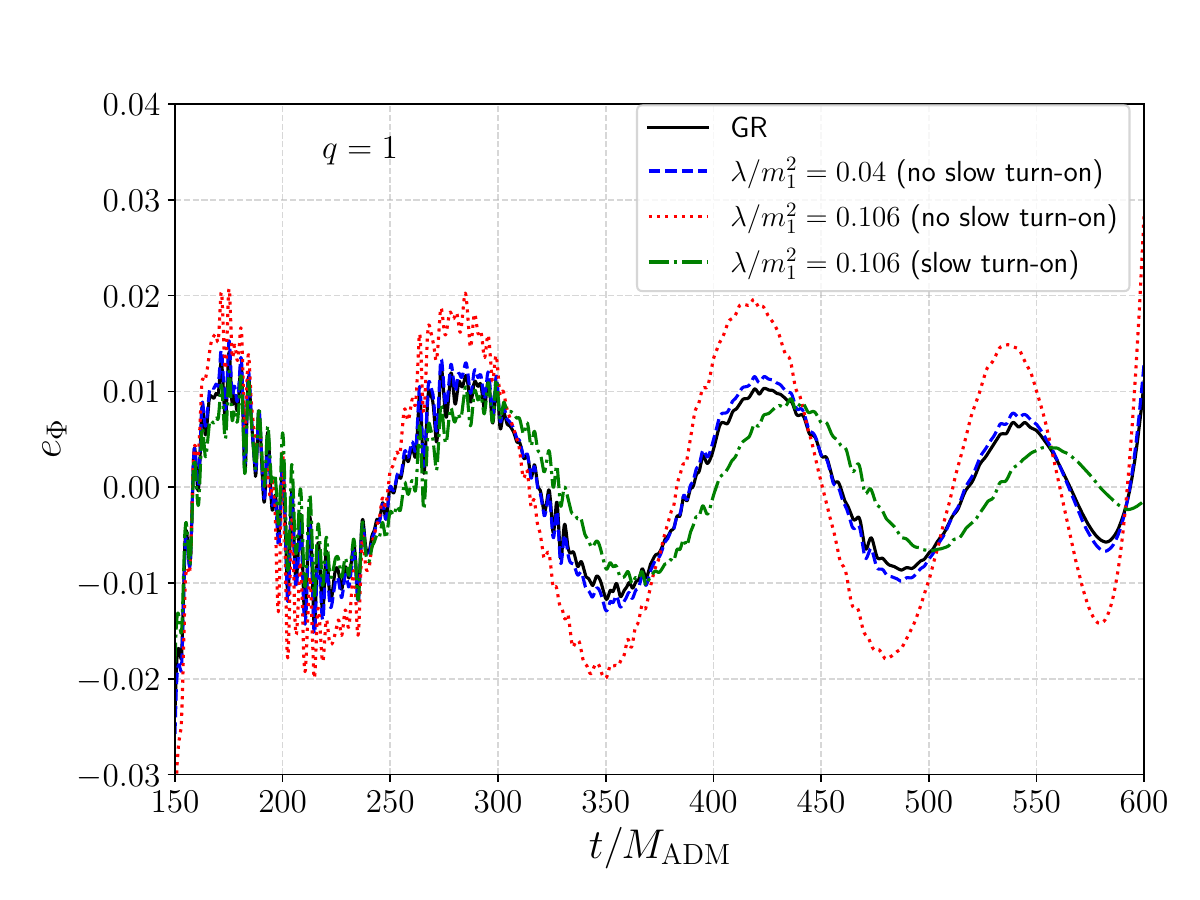}
	\caption{The calculated eccentricity $e_\Phi$ for $q=1$ mass ratio black hole mergers in GR and shift-symmetric scalar-Gauss-Bonnet theory of gravity for GR and $\lambda/m_1^2={0.04, 0.106}$. For some values of $\lambda/m_1^2$ the turn-on of the coupling is immediate, while for others there is a slow turn-on.}
    \label{fig:ecc_q1}
    \end{figure} 
    
    The eccentricity for the $q=1$ mass ratio simulations is presented in Fig. \ref{fig:ecc_q1}. Here, the initial data that we start with have $e_\Phi$ reaching roughly $0.01$. In that case, the eccentricity change for $\lambda/m_1^2=0.106$ with respect to GR is more noticeable. We plot two cases -- with an immediate turn-on of the GB coupling and with a slow turn-on. In the former case, the eccentricity change is larger, almost double the GR value. This appears to be the cause for the larger dephasing with respect to GR observed in Fig. \ref{fig:q1q2} for that particular simulation, whereas the case with a slow turn-on leads to a smaller eccentricity and better match with the PN results.
    
    \bibliography{references}

\end{document}